\documentclass{aa}

\usepackage{txfonts}
\usepackage{graphicx}

\begin{document}

\title{A broadband FFT spectrometer for radio and millimeter astronomy}
\author{A.O. Benz\inst{1}, P.C. Grigis\inst{1}, V. Hungerb\"uhler\inst{2}, H. Meyer\inst{1}, C. Monstein\inst{1}, B. Stuber\inst{3}, and D. Zardet\inst{4}}
\offprints{A.O. Benz, \email{benz@astro.phys.ethz.ch}}
\institute{Institute of Astronomy, ETH, CH-8092 Zurich, Switzerland 
\and Acqiris SA, 1228 Plan-les-Ouates, Switzerland
\and University of Applied Sciences Solothurn, Nordwestschweiz, Switzerland
\and ZMA, Center for Microelectronics, University of Applied Sciences Aargau, Nordwestschweiz, Switzerland
}

\date{Received / Accepted }

\abstract{The core architecture, tests in the lab and first results of a Fast Fourier Transform (FFT) spectrometer are described. It is based on a commercially available fast digital sampler (AC240) with an on-board Field Programmable Gate Array (FPGA). The spectrometer works continuously and has a remarkable total bandwidth of 1 GHz, resolved into 16384 channels. The data is sampled with 8 bits, yielding a dynamic range of 48 dB. An Allan time of more than 2000 s and an SFDR of 37 dB were measured. First light observations with the KOSMA telescope show a perfect spectrum without internal or external spurious signals. 

}

\titlerunning{ Broadband FFT spectrometer}
\authorrunning{A. O. Benz et al.}
\maketitle

\section{Introduction}

Spectrometers from microwaves to sub-millimeter waves are used to detect and to measure molecular lines ubiquitous in star forming regions, planetary and cometary atmospheres. The higher the frequencies, the broader are the lines. Several new instruments are currently being put into operation, such as APEX, Nanten2 and SOFIA, some of them supplied by multiple beams. Multi-feed single dishes for sub-millimeter waves pose a serious demand on the capability to survey simultaneously a sufficiently large spectral band. Such telescopes may well be cost-limited by the spectrometers.

In the past, broadband spectrometers for millimeter waves were mostly based on the acousto-optical design, utilizing the diffraction of coherent light. It was well developed over the past three decades (Cole 1968, Lecacheux et al. 1998, Schieder et al. 2003). However, the development of receivers to higher frequency and larger bandwidth has been even faster. Acousto-optical spectrometers are limited in the number of channels, and pose problems with stability, dynamic range and cost. Digital techniques have also been used in the past, but were even more limited in bandwidth (e.g. Lee et al. 2004; Leitner, Rucker \& Lecacheux 2005). Initially, digital autocorrelation was favored for technical reasons (Weinreb 1963) and was done continuously by hard-wired registers, the Fourier transformation being carried out subsequently. Autocorrelators have reached bandwidths of 500 MHz and 1000 channel resolution in a single device (e.g. Belgazem et al. 2004). Digital spectrometers based on Fourier transformation have been realized with a bandwidth of only 50 MHz and 1024 channels (Stanko, Klein \& Kerp 2005). 

There are many reasons why digital signal processing of an analog signal may be preferable. Most importantly, a digital programmable system allows flexibility. The system can be reconfigured and improved by changing the program; it may be transported at least in part to  new hardware. The stability and accuracy of a digital system is well controlled, once the signal is digitized. In some cases a fully digital spectrometer is cheaper than its analog counterpart. The lower cost may be due to cheaper hardware, or it may be the result of flexibility in the software. Digital techniques promise flexibility and a development synchronous with commercial electronics. 

With the general development of computing power, direct digital Fourier transform becomes competitive with autocorrelation. It is more efficient if under certain conditions the Fast Fourier Transform (FFT) algorithm can be implemented. For the same computing power, it then allows for more channels and higher spectral resolution than autocorrelation. Fourier spectroscopy involves three operations: (i) sampling the intermediate frequency voltage at high resolution, (ii) computing the squared magnitude of the Fourier transform of each windowed data section, and (iii) averaging the spectra. The first step requires digital samplers (A/D converters) with high sampling rate and stability. The second and third steps, however, are even more demanding and constitute the actual bottle neck. The number of computations per second involved in broadband FFT spectrometry (second step) is enormous if combined with the requirement that sampling and computations must be continuous without any loss of data for reasons of sensitivity. Such a demand can be handled only by large programmable gate (logic) arrays. In recent years, Field-Programmable Gate Arrays (FPGA) have been developed with sufficient computational power. These chips consist of up to millions of transistors, supporting ten thousands of logic cells that can be programmed. The active silicon area on the chips has continuously been enlarged, the semiconductor structures have been reduced to a fraction of a micrometer, and the connections built in many layers. The core of an FPGA operates at low voltage, reducing the electric power requirement. 

Here we describe the development of an FFT core and spectrometer with wide bandwidth based on a commercially available sampler with a built-in FPGA chip. The FFT core for the FPGA can be easily down-loaded and controlled from an ordinary personal computer (PC) using the PCI-bus. Thus communications with the spectrometer and data transfer are simple and reliable. The numerical architecture, technical design and development are outlined in Section 2. In Section 3 the first results and experiences from millimeter and sub-millimeter observations at the KOSMA telescope on Gornergrat are described. A summary of the specifications and conclusions are given in Section 4.

\begin{figure}
\centering
\resizebox{\hsize}{!}{\includegraphics{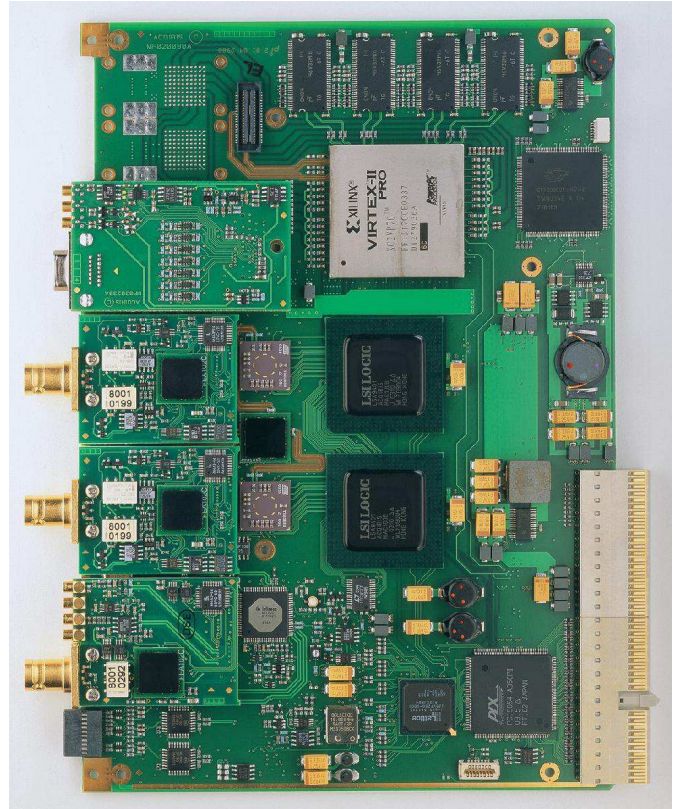}}
\caption{Board of AC240, the hardware of the FFT spectrometer. On the left, from top to bottom, are the system control interfaces, two analog inputs for the signal at intermediate frequency (SMA connectors will be used in the final version), and an input for a trigger or a clock. The FPGA processing unit is visible in the top center, the PCI bus for data output at the bottom right.}
\label{AC240Karte}
\end{figure}

\section{Design and development}

\subsection{Hardware}
The Acqiris signal analyzer AC240 is an 8-bit CompactPCI digitizer, designed to cover requirements encountered in real time frequency analysis applications such as spectroscopy (Fig. 1, Acqiris 2005). The AC240 has two synchronous input channels, each with a 1 Gsample/s A/D converter. The ADCs in the 2 channels can be interleaved and combined to form a single 2 Gsample/s 8-bit converter. The Nyquist theorem states that this time resolution samples a frequency range from 0 to 1 GHz. 

The AC240 timebase controls both acquisition channels simultaneously. The sampling frequencies are set either internally or with an external clock input on the front panel. The choice between the internal and external clock is software programmable. We use a sampling frequency of 2 Gsample/s.

The AC240 board includes a digital processing unit. The on-board FPGA is a Xilinx Virtex II Pro 70 (XC2VP70) with more than 74000 logic cells. It is capable of executing multiplications in less than 5 ns and offers 328 dedicated 18-bit x 18-bit multipliers with 36-bit results, and nearly 7 Mbits of on-chip processing memory. The PCI bus transfers data to a host PC at sustained rates up to 132 Mbyte/s.

With the AC240 board, Acqiris provides a firmware development kit (FDK) that helps to develop custom applications for the on-board FPGA. The FDK consists of a set of standard cores that can be instantiated in the FPGA, mainly interface blocks to the off-chip components. Such interfaces could be the local bus interface (PCI) for connection to a PC, or interfaces for data acquisition, external memories, external DAC, high speed links, micro SUB-D15 or the two front panel MMCX connectors. Embedded in this firmware, we developed a user core in VHDL that fast-Fourier transforms the signal for spectral analysis.

\subsection{FFT core}
There are several approaches to digital spectrometry, including Fourier transformation, autocorrelation and digital filterbanks. For an overview and mathematical derivations, the reader is referred to the textbook by Proakis \& Dimitris (1992). Autocorrelation requires a similar number of multiplications as a discrete Fourier transform (order of $N(N-1)$ where $N$ is the number of data points). However, the FFT needs only 0.5$N$log$_2N$ products if $N$ is a power of 2. Digital filterbanks may be designed to have even less channel crosstalk. They need in most cases about twice the number of multiplications as an FFT. We have decided to use the FFT algorithm.

The signal in the range from initially $f_1$ to $f_2$ is down-converted to the baseband from 0 to $f_2 - f_1$. This intermediate frequency band is sampled periodically at rates of up to 2 Gsample/s. The maximum frequency span, $f_2 - f_1$, is thus 1 GHz. It can be reduced (at an equal number of channels) via software, using a lower sampling rate. To analyze a range from 0 to 1 GHz at intermediate frequency, the signal must be sampled every 0.5 ns. The number of useful frequency channels is half the number of samples per transform. We use a 32768 point FFT, thus yielding 16384 channels. The time to aquire this data is 16.384 $\mu$s. To avoid telescope dead time, the FFT must be computed, squared and the spectra summed within this time.

The sampling time can be increased by a factor of 2 by software, reducing the total bandwidth to 500 MHz. The number of channels remains constant. Further reduction in total bandwidth is possible using the low-frequency option of the FFT core.

The discrete Fourier transform (DFT) is defined by
\begin{equation}
X(k)\ =\ \sum_{n=0}^{N-1}x(n) \exp [-2\pi i nk/N]\ =\ \sum_{n=0}^{N-1}x(n) W_N^{nk}
\label{DFT}
\end{equation}  
where $x(n)$ are the measured samples, and $X(k)$ is the spectrum in each channel $k=0,1,2,...,N-1$. Eq. (1) defines also the twiddle factors $ W_N^{nk}$. The same twiddle factors are used many times in the course of one transformation. They are stored in the FPGA$'$s RAM. The observing frequency $f(k)$ before down-conversion to intermediate frequency is 
\begin{equation}
f = f_1 + {k\over{N \Delta t}}\ \ \ ,
\label{f}
\end{equation}  
at the lower edge of channel $k$, where $\Delta t$ is the sampling time. The spectrum given by Eq. (1) is complex as it contains the information on the phase. The spectral power density is defined as $\mid X(k) \mid^2$.

A computationally efficient algorithm for the DFT decomposes the N-point operation into smaller portions. Eq. (1) is recursively decomposed into a number, $R$, of sequences with equal length. $R$ is called the radix of the FFT computation. This implementation is based on two 32k-FFT pipelines operating in parallel. Complex input data are required to make use of the full capacity of the DFT computation.  For this reason, a method is applied that computes the N-point FFT using an N/2-point algorithm. The input data stream consisting of even and odd samples is treated as a set of complex pairs. An additional processing stage using twiddle factors yields the final N-point result (Proakis \& Dimitris (1992).

\begin{figure*}
\centering
\resizebox{\hsize}{!}{\includegraphics{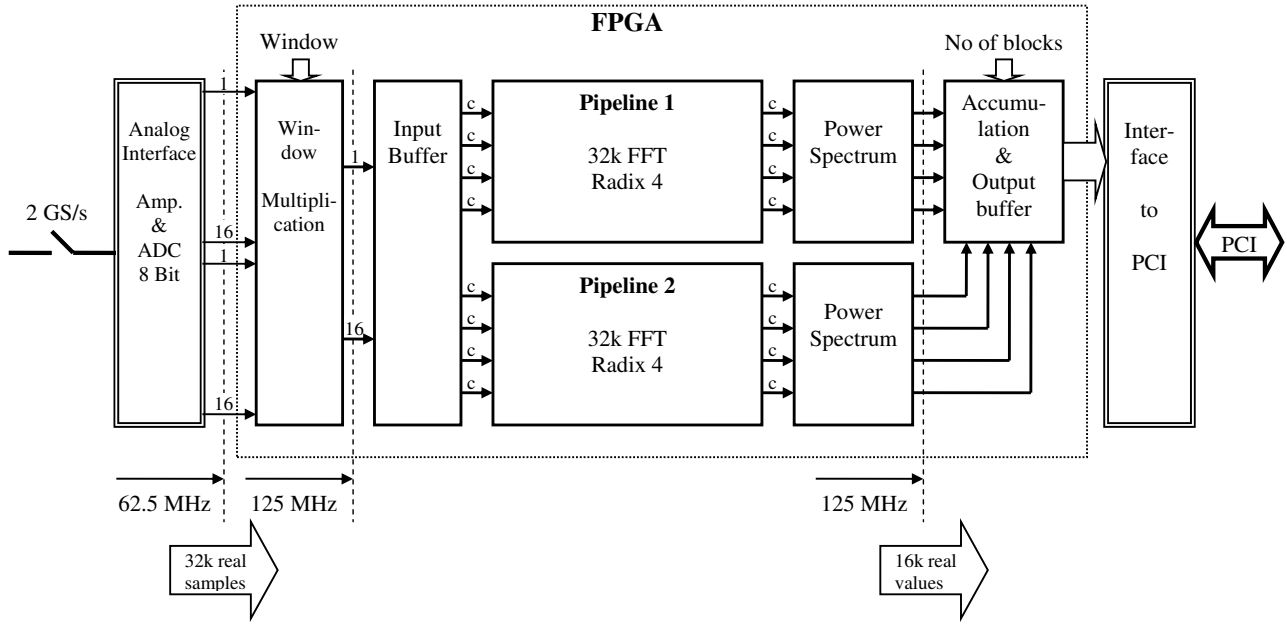}}
\caption{Architecture of the Fast Fourier Transform operations in the FPGA chip of the FFT spectrometer. The letter $c$ indicates that complex numbers are transfered}
\label{block_diagramm}
\end{figure*}

The FFT design in the FPGA is presented in Fig. 2. The A/D converter samples the signal with a resolution of 8 bit. The word width increases during the process to keep the numerical accuracy at a maximum. A windowing function can be applied on the ADC output producing an output of 9 bits. The windowing function is optional. At the moment there are two windows available: boxcar (incl. unfiltered) and Kaiser. The windows are symmetric and 9 bit. After the pipeline the word width is 18 bits, and is enhanced to 34 bits after taking the power spectrum. The PCI bus reads a selectable 32 or all of the 36 bits after accumulating the requested number of spectra in the output buffer. 

The input data throughput of 2 Gsample/s is delivered at a rate of 1 GHz/16 = 62.5 MHz, providing 2$\times$16 words for each clock cycle. Internally the clock is doubled, therefore resulting in 2$\times$4 complex data streams at a rate of 125 MHz. This clock may be stopped for reasons of data acquisition, triggering or synchronization. The synchronization between the pipeline clock and the clock of the PCI bus is achieved with an output buffer using the Xilinx RAM. These RAM blocks are dual-ported with two completely independent ports. 

The PCI transfer time of data output to the PC is 0.5 ms per block (equal to 16'384 data words). It defines the maximum discharge rate for spectra from the FPGA and thus the highest time resolution. Shorter accumulation times and higher time resolution can be achieved only with data gaps.

\subsection{PC interface and data structure}
A regular PC is used to control the AC240 and to receive the data via the PCI interface. Additionally, some control and status signals are available via a connector on the front panel. The present design for interactive control offers 6 bits of digital input and 8 bits of digital output. All bits are available to the user through a standard micro SUB-D15 connector. The input bits can also be read through a register controlled by the PC and the output bits can be written similarly. In the version used for these test observations, the AC240 was used as the client for I/O control, as PC timing was not sufficiently stable. The digital output can nevertheless be used to control the instrument, e.g. switching the telescope position, the beam position, wave polarization, local oscillator frequency (frequency switching) or the calibration unit. In our configuration, the connection to the PC is realized via integrated driver chips. An additional MMCX connector at the front panel of the AC240 (see Fig. 1) offers external clocking or a 10 MHz reference clock for ADC and FPGA hardware. To make use of this external clock or reference clock feature, a different configuration bit has to be set. Standard sampling rates are possible and provided by an internal clock source from 2 Gsample/s down to 1 Gsample/s. An SMA connector is reserved for an external trigger for simple sampler applications. It is not used in our application.

Spectrometer configuration, parameterization and data transfer are accomplished via a PCI-interface card for standard PCs. Industry supports two different interfaces: PXI and CompactPCI. The cheap solution, IC400, uses a single ribbon copper cable 1.2 m long with a maximum transfer rate of 32 bits at 33 MHz (132 Mbytes/s). A combined interface PXI/CompactPCI-express, IC414, is an alternative, offering a RJ45-like cable 5 m long with a maximum transfer rate of 64 bits at 66 MHz (528 Mbytes/s). An important advantage of this second version is the possibility to daisy-chain several crates carrying one or more AC240 boards.

All programmable parameters of the spectrometer are stored in a configuration file that can easily be maintained using a simple text editor. Among others, the following parameters are stored in the configuration file: number of accumulation steps, coupling (50 $\Omega$, dc or ac), selection of anti-aliasing filter width, analog input channel number, input full-scale range and sampling interval. Some special operation parameters can be chosen by selecting appropriate buttons on the desktop window of the PC application.

All acquired spectral data are sequentially stored in FITS format. In our configuration the keywords for the FITS header are either taken from the configuration file, from manual input, or are fixed in software code, depending on the nature of the keyword. All configuration parameters are stored in the header block of each FITS file as a documentation of a particular observation. FITS files are identified by their filenames composed of PC date and time. FITS files can be stored in any storage device connected to the PC. The necessary link is made using a directory keyword in the configuration file. If enabled, all logging information like debug values, error messages, core and board temperatures are also saved in an ASCII coded text file.

The front panel of the AC240 comprises 3 LEDs. Two of them are used to indicate overflow at input or output. 
If the analogue input signal is within the ADC$'$s range, LED L1 is green. If the applied level saturates the ADC, L1 changes to red. LED L2 is green when the accumulation buffer is within its maximum range and changes to red if the accumulation buffer overflows. In the case of ADC saturation, the user or the higher level software can either reduce the input power or, if possible, increase the full-scale range of the ADC by reducing the gain of the input amplifier. If the second LED turns red, one can reduce the input level, increase the full-scale range or reduce the accumulation time. If none of the LEDs are green or red, the spectrometer is in an idle state and no spectra are produced. A third LED, L3, shows the PCI status. It is red while the firmware file is up-loaded.

\begin{table}
\begin{center}
\begin{tabular}{lr}   
\hline \hline
Parameter                          & result, unit        \\
\hline%
Input sampling rate max            & 2 Gsample/s     \\
Acquisition time per spectrum min  & 16.384 $\mu$s    \\
FFT execution time per spectrum    & 16.384 $\mu$s      \\
Observation bandwidth max          & 1 GHz         \\
Observation bandwidth min          & dc         \\
Input resolution ADC               & 8 bit              \\
Input full scale range             & 50 mV...5 V         \\
Input coupling                     & ac~or~dc~50$ \Omega$ \\
Maximum dynamic range              & 48 dB              \\
Ghost-free dynamic range           & $>$ 30 dB   \\
Spurious-free dynamic range SFDR   & 37 dB              \\
Intercept point IP2 (5V range)     & +1 dBm             \\
Intercept point IP3 (5V range)     & -17 dBm            \\
Adjacent channel rejection         & 19 dB          \\
Data width (FFT pipeline)          & 18 bit             \\
Output digits via PCI 33 MHz       & 36 bit             \\
Data transfer time                 & 0.5 ms           \\
Windowing                          & programmable      \\
Number of channels                 & $4^{7}$ = 16384    \\
Channel separation                 & 61.035 kHz         \\
Reception bandwidth -3 dB (boxcar) & 54 kHz            \\
Reception bandwidth -10 dB (boxcar)& 90 kHz             \\
Allan time in the lab ($\pm$ 0.2K) & $>$ 2000 s            \\
Accumulation on board min          & 32.768 $\mu$s    \\
Accumulation on board max          & $1.4\cdot10^5$s$\sim 39$h   \\
Electrical power consumption	     & $<$ 100 W              \\
Weight without PC                  & 7.8 kg             \\
Operating temperature              & $0^{\circ}...40^{\circ}C$\\
Relative humidity (non condensing) & 5...95\%          \\
Dimensions (crate CC103)           & $342 \times 346 \times 106$ mm \\
Shock (half sine pulse)            & 30 G               \\
Vibration (random)                 & 5...500 Hz         \\
EMC immunity compliant to          & EM61326-1         \\
EMC emissions compliant to         & EM61326-1         \\
 \hline
\end{tabular}
\end{center}
\caption{Specifications and performance parameters measured in the lab or at KOSMA observatory (Gornergrat).} \label{performance}
\end{table}

\section{Observations and Results}

\begin{figure}
\centering
\resizebox{\hsize}{!}{\includegraphics{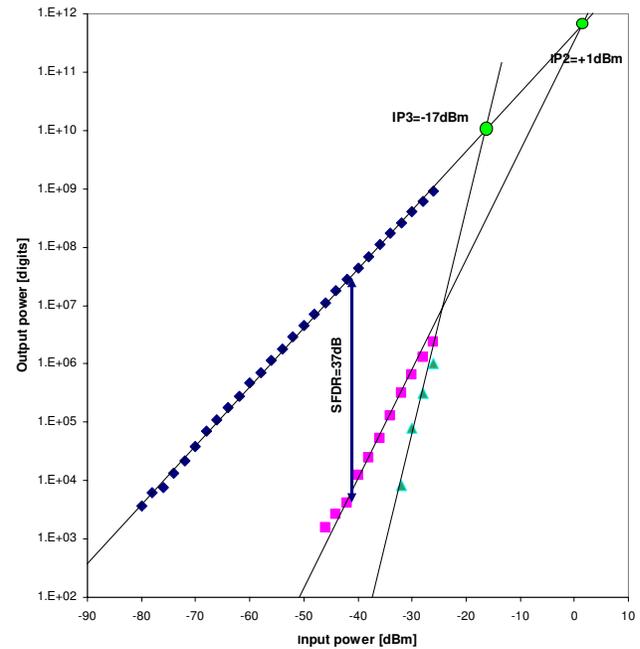}}
\caption{Uncalibrated units of output power vs. input power of a narrowband signal generator. The input of broadband background noise corresponds to $10^3$ digits (uncalibrated units). The power of the signal at 305.17 MHz (channel 5000) is shown with black diamonds. The measured output power at the first harmonic (channel 10000) is represented with gray squares and the power at the second harmonic (channel 15000) with gray triangles. The intersection of the extrapolated regression lines are defined as intercept points given in Table 1 and discussed in the text.} 
\label{SFDR}
\end{figure}

\subsection{General performance}
The FFT spectrometer was tested with nearly final hardware and firmware, and a development version of the FFT core and PC interface. In this subsection some of the more important laboratory tests are described. For Fig. 3 a narrowband signal (much narrower than the channel width) was fed into the FFT spectrometer. To avoid digitization errors in the other channels, the signal of a broadband noise generator was also applied. It produced a background level of $10^3$ units. Note the linearity of the FFT spectrometer in the relation of output power vs. input power. Deviations from a straight line are visible at the top, where the 8-bit A/D conversion approaches saturation (theoretical limit is 48 dB), and in the first harmonic at the bottom, where some influence of the background becomes noticeable. 

The spurious free dynamic range (SFDR) specifies the quality of the amplifier and A/D converter. The SFDR is the ratio of the amplitude of the applied signal to the value of the largest harmonic. From Fig. 3 in double-logarithmic representation, the SFDR is found at 37 dB. Figure 3 also shows the extrapolated intercept points, +1 dBm and -17 dBm, where the second and third harmonics, respectively, would become equal to output power of the applied signal. 

\begin{figure}
\centering
\resizebox{\hsize}{!}{\includegraphics{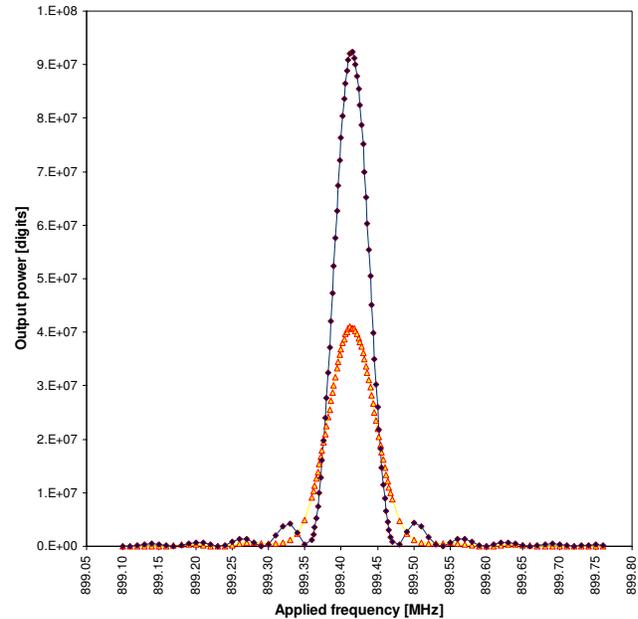}}
\caption{Filter curve obtained by measuring the output of frequency channel 14736 (899.414 MHz) while a signal generator producing a very narrowband signal at the applied frequency is stepped through the band. The observed output power of the FFT spectrometer in unfiltered mode, corresponding to a boxcar filter with one channel$'$s width, is given with diamonds, the result applying a Kaiser filter is displayed with triangles.}
\label{Filter}
\end{figure}

The spectral resolution and effective channel bandwidth of the spectrometer was investigated from the filter curve in Fig. 4. It is the output power in the channel centered at 899.414 MHz resulting from a narrowband signal (much narrower than the channel width) that was moved in frequency from 899.050 GHz to 899.800 GHz. The frequency step varied from 10 kHz far from the observing channel to 0.2 kHz near its center. The result shown in Fig. 4 has the shape of a sin$x/x$ curve, as predicted from the discrete response to a delta function in frequency. The main peak is exactly in the center of the chosen bin. The secondary peaks are damped according to the equation $20 \log_{10}(k \pi/2)$, where $k$ = 3, 5, 7, ... and $k$ = 3 is the first secondary peak. From the filter response curve in Fig. 4, a reception bandwidth of 54 kHz at -3 dB and of 90 kHz at -10 dB is derived for a boxcar window equal to the channel width (unfiltered). When the signal is in the middle of the center channel, the output power measured in the two adjacent channels is 19 dB below the power in the center channel.

\begin{figure}
\centering
\resizebox{\hsize}{!}{\includegraphics{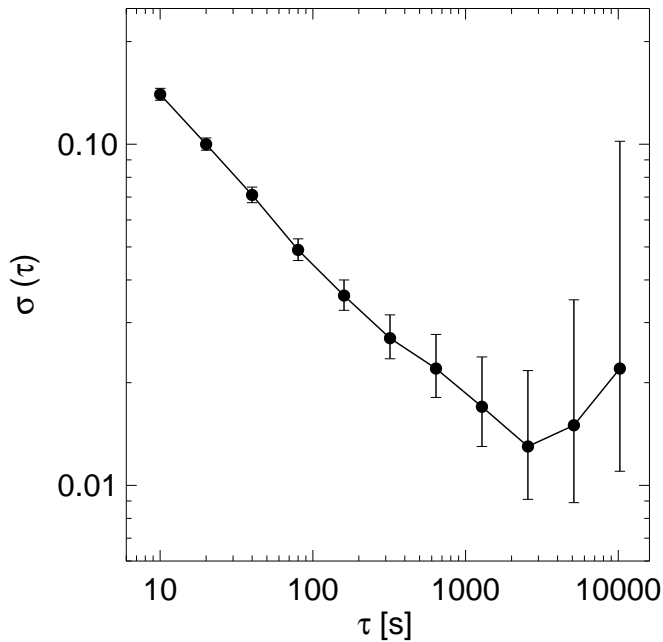}}
\caption{The variance $\sigma$ of a frequency channel vs. integration time $\tau$. The Allan time is defined by the minimum in $\sigma$.}
\label{Allan}
\end{figure}
The Allan variance was measured with a 50 Ohm load resistor in an air conditioned room with temperature variations $\pm$ 0.2 K, resulting in an Allan time $>$ 2000 s (Fig. 5). When the temperature stabilization was reduced, allowing larger variations, the Allan time diminished accordingly until, at more than $\pm$ 2K, it fell to $\approx$ 200 s. This proves that the stability of the FFT spectrometer is determined by temperature variations either of the load resistor or in the analogue electronics of the ADC. 

The frequency of 62.5 MHz, the pipeline clock frequency, and most of its harmonics have an enhanced background. Surprisingly, the internal spurious signals were noticeably reduced when a slot PC 503 was used. The internal interferences disappear after background subtraction, thus seem to be stable. We noted however that the frequency channels of 62.5 MHz and the harmonics generally have shorter Allan times (by about one order of magnitude), which improved with temperature stability. A different kind of spurious signal was visible at the frequency emitted by a TV station near the lab. The PC clock frequency, however, did not produce observable interference. Finally we searched for ghost images of signals, produced by slight mismatches in phase and gain of the two interleaved ADCs. When a signal is introduced at a frequency $f_1 + f_S$, a ghost signal may appear at $f_2 - f_S$. Ghost images of strong radio interference have been noticed in baseband observations of the radio sky. The strength of the ghosts was measured at 28 - 40 dB below the signal, defining a 'ghost-free dynamic range' of about 30 dB or more. Its high value indicates that the hardware is well tuned.

The maximum power consumption of the AC240 board alone is 73 W. The spectrometer including the power supply unit without the PC consumes less than 100 W, and 66 W in idle mode.

\begin{figure}
\centering
\resizebox{\hsize}{!}{\includegraphics{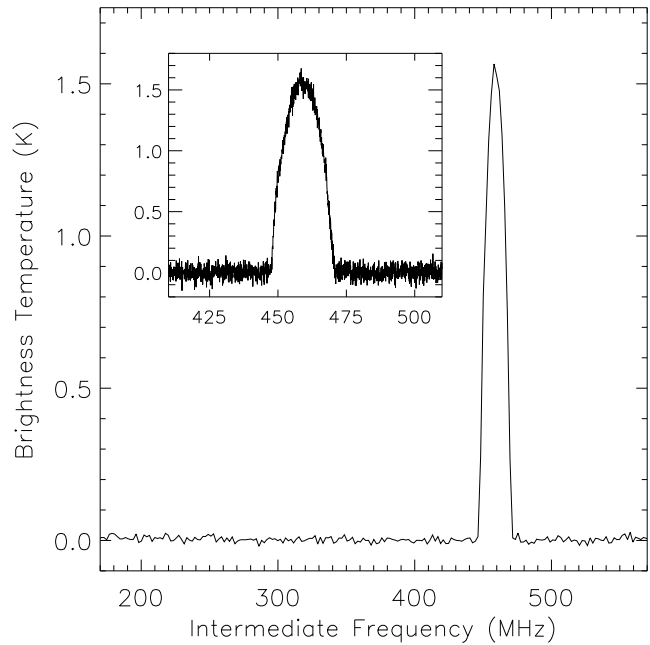}}
\caption{{\it First light} spectrum around the relatively weak CO(2-1) line of the evolved carbon star IRC +10216 observed with the FFT spectrometer for a net on-source time of 57 minutes. The frequency is the intermediate frequency as measured by the spectrometer. To reduce noise, 32 channels are integrated (resolution 1.95 MHz). The insert is an enlargement of the line in full resolution (61 kHz). }
\label{IRC}
\end{figure}

\subsection{Observations of celestial sources}
As the FFT spectrometer is easily transportable, it was brought to the KOSMA observatory at Gornergrat (Zermatt). The diameter of the telescope there is 3 m, and the available receiver had a total bandwidth of about 300 MHz, which was converted down to 200 - 500 MHz. First light was observed on March 21, 2005. Figure 6 shows a high signal-to-noise spectrum over the available band. The off-source emission was subtracted. The data was calibrated using an existing spectrometer of the observatory, operating in parallel. Figure 6 indicates that the spurious signals caused by harmonics of the clock frequency, for example at 437.5 MHz and 500 MHz (insert), do not leave any traces in real observations. Radio interference from external sources at Gornergrat was not visible either. No ghost image of the line, mirrored at 500 MHz, is noticeable.

\begin{figure*}
\centering
\resizebox{\hsize}{!}{\includegraphics{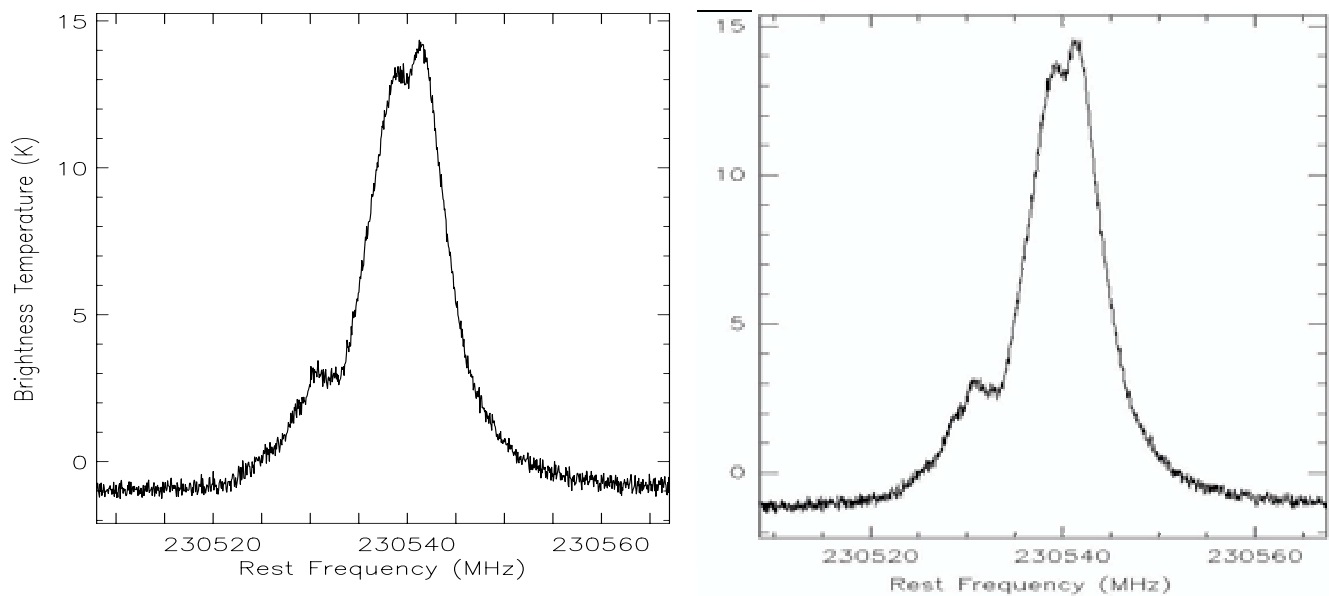}}
\caption{{\it Left:} The CO(1-2) line of the high-mass protostar DR 21K, observed with the FFT spectrometer for a net on-source time of 90 s. {\it Right:} A comparable observation with the high-resolution acousto-optical spectrometer of KOSMA.}
\label{DR21K}
\end{figure*}
In Fig. 7, the FFT spectrometer is compared in a fraction of its frequency range with a narrowband, high-resolution acousto-optical spectrometer on site. The FFT spectrometer observed simultaneously for about half of the time at a similar reception bandwidth. The noise of the two observations is comparable, consistent with the expectation that it does not originate from the spectrometers.

\section{Conclusions}

We present here the architecture of a fully digital spectrometer based on the FFT principle. For the first time the FFT analysis is realized for a broad bandwidth. In number of channels, 16384, the new broadband FFT spectrometer exceeds the existing techniques (acousto-optical and autocorrelation) by an order of magnitude. Specifications and performance parameters measured in the lab or at the KOSMA observatory are listed in Table 1. The design is aimed at a low-cost module that can also be used in multi-device applications. The hardware is commercially produced and available 'off the shelf' together with the FFT core developed in this project. The PC software, user interface and data handling software are public. A simple one-device version has been tested, allowing the validation of the FFT approach and the architecture. The tests have shown that the fully digital FFT spectrometer is not only a great enhancement in technical performance, but improves availability and user-friendliness. 

The 8-bit sampling is sufficient for millimeter and submillimeter astronomy, but may pose a problem at lower frequencies when terrestrial interference exceeds the noise by more than 48 dB. Nevertheless, the described spectrometer is a remarkable improvement in view of the more than 20 dB smaller dynamic range of present acousto-optical spectrometers. A future extension to a 12-bit A/D conversion would solve most problems in low-frequency FFT spectroscopy. Considering the demand for broad bandwidth for sub-millimeter spectroscopy, the development of samplers with higher rates seems more pressing. The development is, however, not limited by the sampling rate, but by the capacity on the FPGA for on-line FFT analysis. Thus, progress by a similar step in bandwidth and channel number cannot be expected in the near future.

\begin{acknowledgements}
M. Pichler helped with the implementation of the FFT in the FPGA environment. We also thank the KOSMA observatory at Gornergrat for the possibility of test observations at short notice and, in particular, Dr. M. Miller, and M. Battaglia for help with observations. K.U. Schenk, B. Klein and Dr. R.T. Schieder are acknowledged for helpful discussions. The technical development was financed by the Swiss Kommission f\"ur Technologie und Innovation (KTI grant nr. 6421.2 IWS-IW). The previous development of a preliminary breadboard model was supported by the Swiss National Science Foundation (grant. Nr. 2000-061559).
\end{acknowledgements}


\end{document}